# Surface quasi periodic and random structures based on nanomotor lithography for light trapping

Sh. Golghasemi Sorkhabi[1,2,a*], S. Ahmadi-Kandjani,[1,a] F. Cousseau[2,b], M. Loumaigne[2,b], S. Zielinska[3,c], E. Ortyl[3,c] and R.Barille[2,b]

[1] *Research Institute for Applied Physics and Astronomy (RIAPA), University of Tabriz, 51666, Tabriz, (Iran).*

[2] *Université d'Angers/UMR CNRS 6200, MOLTECH-Anjou, 2 bd Lavoisier, 49045 Angers, (France)*

[3] *Department of Polymer Engineering and Technology, Wroclaw University of Technology, Faculty of Chemistry, 50-370 Wroclaw, (Poland)*

We compare the characteristics of two types of patterns obtained with two azopolymer materials: a Gaussian random pattern and a quasi-random grating pattern. The surface structurations have been obtained with a simple bottom-up technique by illuminating azopolymer thin films with a single laser beam. We demonstrate the interesting generated properties of these two surfaces. In particular, the surface with quasi-random gratings can address beam splittings for light coupling in different directions in an ultra-thin film. We use these two surfaces as a mold and replicate them on a transparent elastomeric material and we demonstrate a very good light entrapment. We also show that the efficiency of light trapping is 20% better with the quasi-random gratings than with the Gaussian random surface, and is close to 40%.

## 1. Introduction

Disordered photonics has experienced a burst of activities in the last two decades. Fabrication of nanostructured materials for control of light-matter interactions has become particularly important for improving different phenomena involving a high degree of light scattering [1]. For example, light emitting diodes necessitate out-coupling of light with a large directivity [2]. The development of solar cells requires the reduction of the active layer. In the latter case, the need for an increment of the optical absorption in order to obtain good performances, can be managed by the use of optimized rough surfaces to achieve a spatial and spectral broad band light trapping. Quasi-random or Gaussian random structures have been proposed as good candidates for energy harvesting.

However, it is still an inquiry if random or periodic photonic nanostructures lead to a better light trapping, and how quickly they could be fabricated [3]. This question has been debated and so far has not found an answer. It remains unclear if solar cells prepared on periodic or randomly textured surfaces exhibit higher efficiencies. Nevertheless, it is generally agreed that diffraction has a higher potential for enhancing the internal path length of light inside a solar cell.

Different methods for material or surface engineering have been proposed with a photon control, based on nanowires, [4] photonic crystal architectures [5], deep reactive ion etching, chemical vapour deposition and electron lithography [6]. All these methods are costly or, depend on required parameters, need multi-step processes. However, methods based on a surface structuration are simpler and more robust, since it is a bottom-up approach. Recently, a lithographic method has been proposed based on the speckle lithography to fabricate 2D random and quasi-random structures. The proposed method works via the generation of a speckle pattern and the projection of this pattern onto a substrate coated with a photoresist [7]. Another method uses an individual speckle diffraction phenomenon for the 1D or 2D random grating fabrication [8]. The diffraction pattern of each speckle forms a micron or sub-micron size grating on the photoresist after exposure and development. In this case, even if the size of micro-gratings could be tuned due to limitations of the speckle method, the surface is not fully covered with gratings.

Here, we propose a new method to generate a 2D quasi Gaussian surface and a quasi-random periodic surface (quasi-random gratings) based on a molecular nano-motor lithography, where the nanoscale movements of molecules acting as nano-motors are fueled by the laser light. This movement can induce self-organized randomly dispersed nanopatterns on the surface of a thin film. The nanomotors are azobenzene molecules. Azobenzene derivatives are one of the most frequently studied classes of switchable compounds. The azobenzene functionality exhibits a *cis-trans* isomerization. The thermodynamically stable *trans* isomer can be converted to the cis isomer by a light stimulus, whose the wavelength is in the absorption band of azobenzene, with a back isomerization induced by a thermal relaxation. Engineering azopolymer materials can lead to a photoisomerization induced by the light, to create self-induced patterns on the surface of thin films. One example of such induced patterns is the production of self-organized structures leading to gratings on the surface of side-chain azobenzene graphted copolymer DR1 thin films [9]. This study was the first experimental result of photoinduced spontaneous structures of azopolymer, generated with a single beam. As such, the use of azopolymers is a good choice for the fabrication of random nanostructured materials for the purpose of controlled light-matter interactions, particularly for the improvement of high light scattering through different phenomena involving in the process. Through this process, different self-structured surface patterns were observed, in particular with epoxy-based azopolymers [10], acrylic polymers bearing photoresponsive moieties as side chains of the polymeric backbone [11] or 4VP(OH-DMA) [12], but none has ever been used for the light management in photonics.

In this study, we compare two quasi random nano-patterned surfaces of azopolymer thin films, capable of being used in the fabrication of devices requiring a light harvesting. We discuss the advantages of both of these surfaces for energy conversion and present two examples of applications requiring a high light coupling in a thin film. As a goal for further developments of light scattering applications, randomly textured surface of the azopolymer materials, prepared through a self-organization process, is used as a master for replication on an elastomeric material. Manufacturing costs of this material are significantly lower compared to random surfaces prepared via lithography or sputtering processes. Moreover, the fabrication of the azopolymer master sample with different patterns can be repeated by erasing the initial surface pattern and exposing it to different light conditions in order to generate tunable patterns.

## 2. MATERIALS AND METHOD

### 2.1 Azopolymer Materials

The first material is a chiral azopolymer (figure 1a), that was synthesized by polymerization of three-component mixtures of azobenzene group containing N-[4-[(E)-[4-[bis(2-hydroxyethyl) amino] phenyl] azo]phenyl]sulfonyl benzamide, chiral 2,3-O-benzylidene-D-threitol and isophorone-diisocyanate (molar ratio of azobenzene derivative, chiraldiol and diisocyanatein that mixture was 1:1:2). The glass transition temperature of the azopolymer 1, determined by DSC (Electrothermal 9100), is 88 °C, and the weight of the average molecular mass of this polymer is 18 400 g/mol.

The second chiral azopolymer (figure 1b) was synthesized by a radical copolymerization of an equimolar mixture of a photoisomerizable azobenzene group containing 2-[4-[(E)-[4-(acetylsulfamoyl)phenyl]azo]-N-methyl-anilino]ethyl 2-methylprop-2-enoate and chiral 2-methyl-N-[(1S)-1-phenylethyl]prop-2-enamide. The glass transition temperature, determined by DSC and the weight of average molecular mass for this copolymer is 77 °C and 14 700 g/mol, respectively.

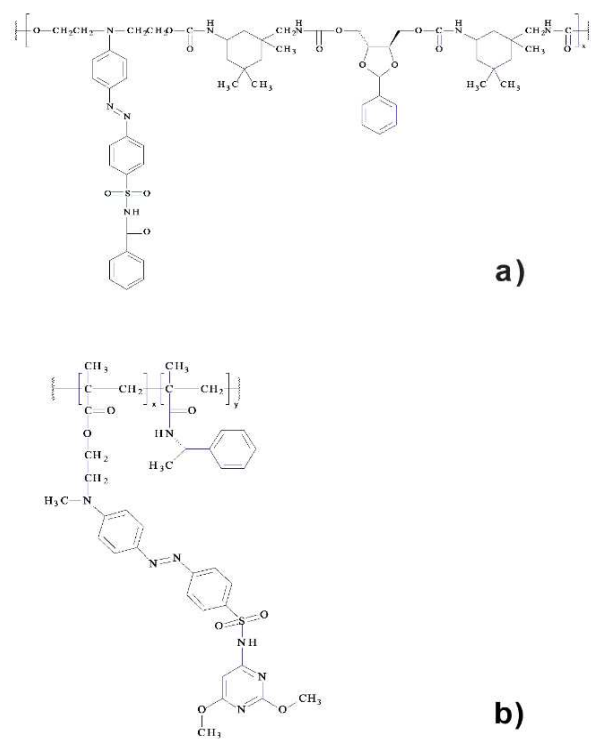

Fig.1. a) Chemical structure of chiral azopolymer1, b) Chemical structure of chiral azopolymer 2.

The two samples have almost the same optical absorption (less than 20% change) and optical spectra (maximum absorption band at $\lambda = 445$ nm). The samples were prepared via a spin-coating technique. The same parameters were used in the spin-coating process to fabricate the samples, and the thicknesses of the different tested samples are comparable and are in the range of 460 – 510 nm (measured with a Dektak 32 profilometer). The chiro-optical properties of the two azopolymers were investigated by circular dichroism spectroscopy and the result confirms the presence of chiral sub-units in the synthesized materials. Some properties of the chromophores and azopolymers were measured with quantum chemical calculations. The geometry optimization of the structures of both azopolymers was optimized using the Gaussian09 package [a] with a RHF method and a 3-21g basis set. The dipole moment and polarizability values of both model structures of chromophores were similar.

## 2.2 Photo-inscription of random surface

A diode pump solid state laser, operating at $\lambda= 473$ nm, is used to excite the azopolymer close to its maximum absorption wavelength. The absorbance at the working wavelength of 473 nm is 0.638. The incoming light intensity is controlled by a combination of a half wave plate and a polarizer. The sample is set perpendicular to the incident laser beam. The size of the collimated laser beam impinging on the polymer sample is adjusted with a Kepler-type afocal system and the beam is a plane wave. The sample is irradiated by a beam with a diameter size of 3 mm at $1/e^2$. The power density is 0.7 W/cm$^2$. A combination of lenses is used to measure the first order diffraction intensity, which allows the evaluation of the photoinduced structuration on the surface of the sample and the molecular rearrangement leading to a mass transport. The sample is illuminated during 15 minutes, until the complete structuration of the thin film surface, which is determined by the saturation of the diffraction intensity.

### 2.3 Random Gaussian surface simulation

The simulated surface is characterized using terms from the probability theory such as: the height distribution function (hdf) or the statistical moments of mean and variance (or rms height [σ]). The surface variations in the lateral directions are described by the auto covariance function (acf), which describes the covariance (correlation) of the surface with translationally shifted versions. The correlation length (τ) is the typical distance between two similar features (e.g. hills or valleys) [13]. The simulation of the random rough surface with a Gaussian statistics is done using a method outlined in [14], where an uncorrelated distribution of surface points, using a random number generator (i.e. white noise) is convolved with a Gaussian filter to obtain the correlation. This convolution is more efficiently performed using the discrete Fast Fourier Transform (FFT) algorithm. Basically, the simulated surface can be generated by [15]

$$z_{p.q} = \sum_{k=0}^{M-1} \sum_{l=0}^{N-1} h_{k\ell}\, \eta_{p+k,q+\ell} \tag{1}$$

With p = 0,1,2,…,(M-1), q = 0,1,2,…,(N-1). $h_{k\ell}$, called the FIR filter (Gaussian filter), is the coefficient defining the system and η is a serie of Gaussian random numbers. Thus, a M x N point surface with a given m x n ACF is generated. m and n should be less than M and N. We take $h_{r,s}$ and $H_{k,\ell}$ as:

$$H_{k,\ell} = \sum_{r=0}^{m-1} \sum_{s=0}^{n-1} h_{r,s}\, exp\left(i2\pi \left[\frac{kr}{m} + \frac{ls}{n}\right]\right) \tag{2}$$

$$h_{r,s} = \frac{1}{mn}\sum_{k=0}^{m-1} \sum_{\ell=0}^{n-1} H_{k,\ell}\, exp\left(-i2\pi \left[\frac{kr}{m} + \frac{ls}{n}\right]\right) \tag{3}$$

With $k_r$ = 0, 1, 2, …., (m-1) and $\ell_s$ = 0, 1, 2, …, (n-1). The FFT of equation1 gives:

$$Z_{k,\ell}(\omega) = H_{k,\ell}(\omega)A_{k,\ell}(\omega) \tag{4}$$

Where, $A_{k,\ell}(\omega)$ is the FFT of η and $H_{k,\ell}$ is the transfer function. We take the inverse FFT of $Z_{k,\ell}(\omega)$ to obtain the simulated random surface.

### 2.4 Extinction measurements

The extinction measurements of the pattern replica were done with a collimated halogen lamp, covering the spectrum from 400 nm to 800 nm, transmitted through the sample. The light is collected with a fiber and a collimating lens and sent to a spectrometer (Edmunds optics).

### 2.5 PDMS replica of azopolymer surface

The Polydimethylsiloxane (PDMS) prepolymer was purchased from Momentive (RTV 615). PDMS is a polymeric organosilicon compound. The PDMS prepolymer was prepared by mixing the elastomer base and the curing agent in a proper ratio (10:1, wt/wt).

A 10 x 10 x 2 mm 3D printed frame was put on top of the thin film with the surface structure and the prepolymer was poured into the frame and dried at room temperature for 2 days.

## 3. RESULTS AND DISCUSSION

### 3.1 Quasi Gaussian surface

The first experiment concerning surface patterns is done by illumination of the azopolymer 1. The figure 2a presents the surface topography of the azopolymer 1 obtained with an atomic force microscope (AFM). The topography of the first sample is a quasi-random pattern. The illumination by a laser light induces a spontaneous self-structured surface pattern with spontaneous created cells, with diameters in the range of nanometer scale giving a granular aspect.

The cell size has an average perimeter in the order of 2.3 µm. The figure 2b gives the Fourier transform of the surface. The lattice has a Fourier energy concentrated into a well-defined circular range of k-vectors with a radius value around 15.2 µm$^{-1}$.

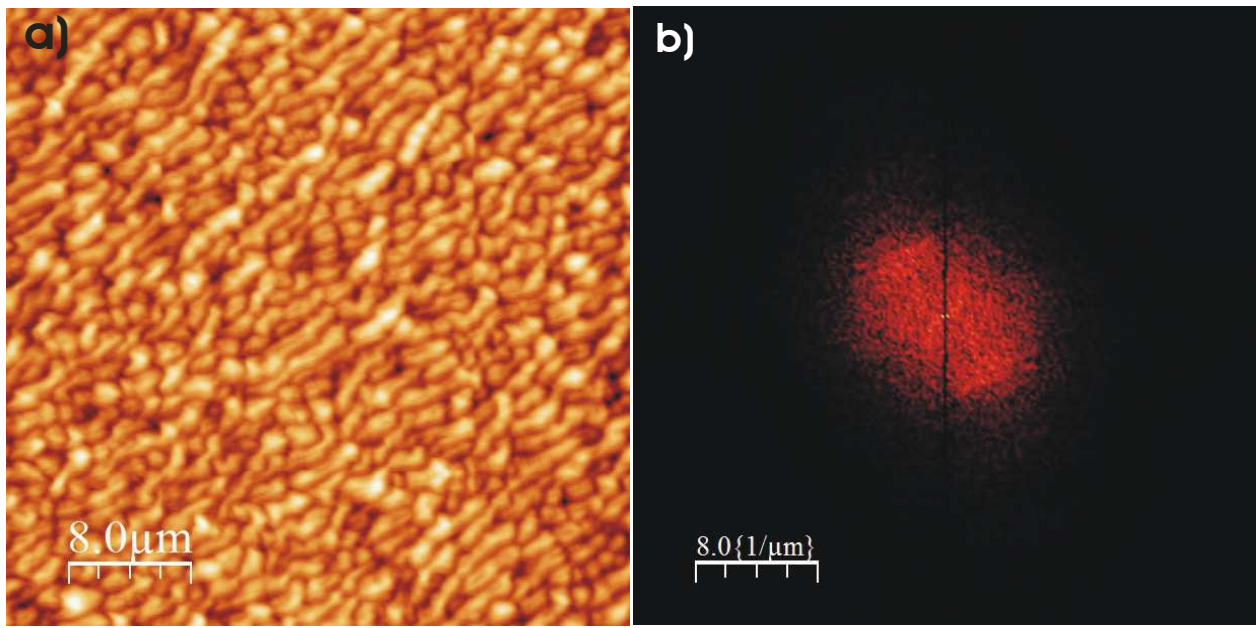

Fig. 2. a) Surface pattern of the azopolymer 1 measured with an AFM in the contact mode, b) Fourier transform of the surface pattern.

A histogram of data given by a statistical roughness analysis, calculated from the topographic measurements acquired by the AFM, is shown in the figure 3a. A fit of the roughness histogram with a Gaussian curve gives an excellent match, and the measured average height is 265 ± 10 nm. The granular pattern has a very small structuration anisotropy, confirmed by the symmetry of the Fourier transform. While the laser polarization was linear during the illumination, no specific correlation of the polarization with the direction of the pattern is observed.

The RMS roughness (σ) is 82 ± 2 nm. In the figure 3b, the correlation function of surface heights is also quite close to a Gaussian curve. We measured the lateral correlation length considering a Gaussian fit (defined as the $e^{-1}$ half-width of the correlation) and we obtained a value of ξ = 510 ± 10 nm. The ratio σ/ξ is 0.16. With this small ratio, slopes of the surface roughness are small. This result is confirmed by the Kurtosis parameter (K), which is a dimensionless quantity giving a measurement of the sharpness of the height distribution function. We found K = 2.7, close to the value 3, for a Gaussian height distribution.

This low ratio of the surface roughness implies that if we consider a ray-tracing, following the geometrical optics, multiple scattering or shadowing effects will be attenuated and will be insignificant at small angles of light scattering (referred to the normal axis). Second order reflections can take place only for high angles of incidence.

The roughness measurements have been done on several spots on one spin-coated sample and confirmed for different samples. The difference between all the measurements differs only by few percents meaning that the spontaneous photoinduced surface patterns are reproducible with the same conditions of solvent and spin-coating parameters.

The surface can be described, in details, with a height–height correlation function (HHCF), H(r) (figure 3c).The measured surface is explained in terms of scaling exponents. The HHCF data reach a saturation height, which can be related to the roughness of the image as well as the average grain size, by utilizing the Hurst function [16] given by the following equation

$$H(r) = 2\sigma^2 \left\{1 - \exp\left(-\left(\frac{r}{\xi}\right)^{2\alpha}\right)\right\} \qquad (5)$$

Where, α is the Hurst parameter (or surface roughness exponent), σ is the surface roughness and ξ is the correlation length. The surface morphology is completely characterized by these parameters. The curve reaches a plateau and the Hurst parameter (α) or surface roughness exponent influences the slope of the function before it reaches the saturation height. The surface roughness is a measure of changes in height across the surface of sample. The correlation length influences the saturation height, which is defined as the largest distance in which the height is still correlated. The surface roughness exponent, α, allows the determination of the fractal dimension and the frequency of height fluctuations. The fractal dimension, $D_f$, corresponds to changes in the surface morphology, which occur due to the effects of the photoinduced surface patterning. The surface roughness is linked to the fractal dimension, $D_f$, of the random surface by $D_f = D - \alpha$, where D is the Euclidean dimension of the surface, (D=3) that defines the geometry of the sample. We fitted the Hurst function, plotted in the figure 3c, to the calculated HHCF data in order to obtain the values of parameters in the equation (5).

The fit shows a great agreement with a model of a Gaussian surface. The surface roughness exponent, α, obtained by fitting the HHCF data of the AFM surface topography is 0.92. The Hurst parameter, α, is close to 1, meaning that the surface is smooth. The fractal dimension, $D_f$, is 2.08.

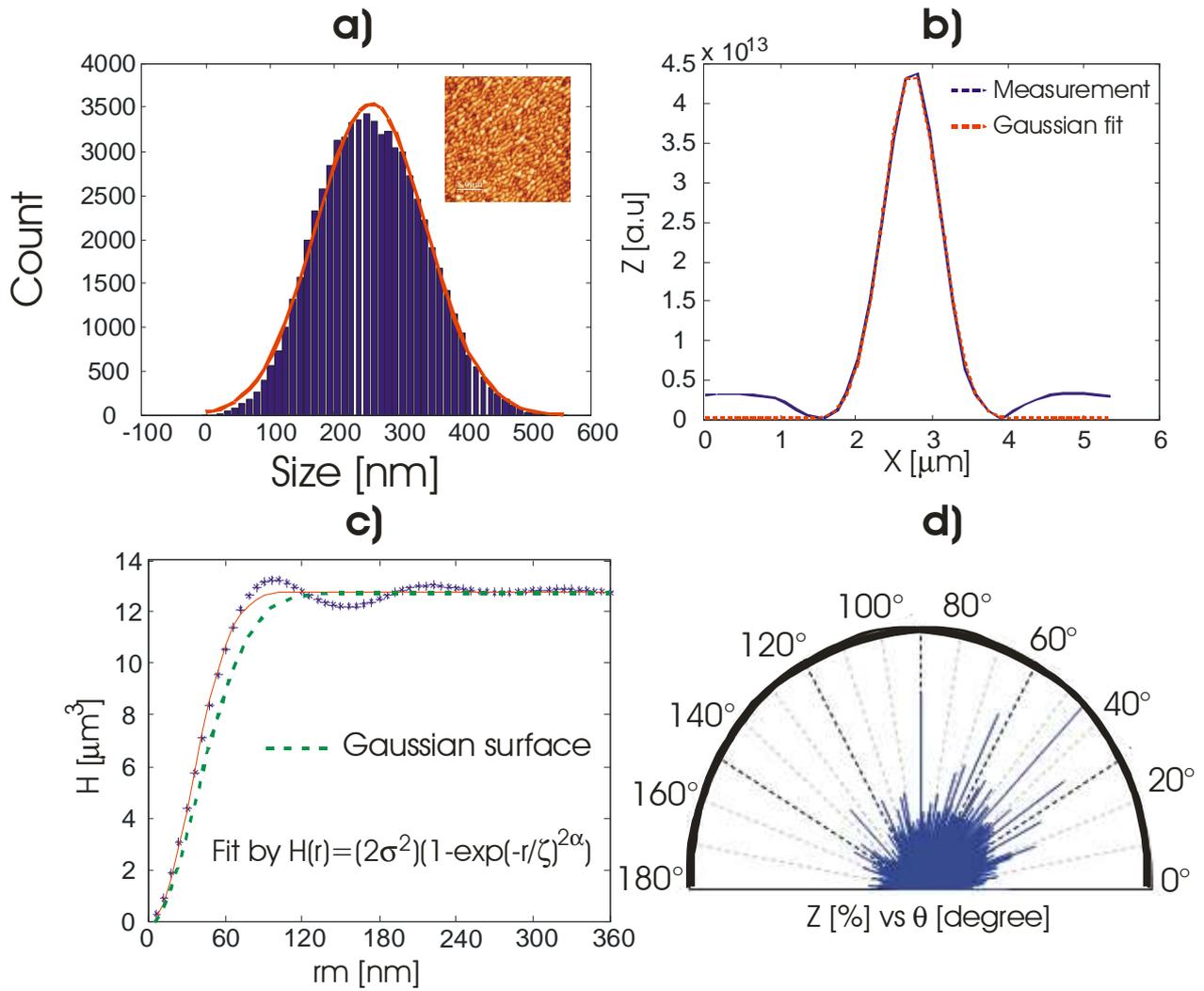

Fig.3. a) Histogram of the surface RMS roughness of the azopolymer 1 and its fit with a Gaussian curve, b) auto-correlation function of the pattern surface and comparison with a pure Gaussian surface (red), c) height–height correlation function (HHCF) and fit, d) directionality of the light emission from the surface.

The directionality of the emission in the figure 3d gives a broadband range of angles. The emission isotropy for this random surface is 52.6 %. This effect is mainly due to the low value of the roughness angles and the smoothness of the surface, as it was measured with the Hurst parameter and the ratio $\sigma/\xi$. This directionality leads to an emission of light into all directions of a hemisphere, with almost a constant radiance, and consequently can be considered as quasi Lambertian.

In order to confirm that the photoinduced surface on the thin film is a random surface, obtained with a stochastic process and resulting from an acting random molecular surface process, we simulated the Gaussian surface pattern, considering the same lateral correlation length as the experimental measurements.

The result of the simulations is given in figure 4. It is seen that we are unable to find information of short range or very short range periodicities in the simulated surface structure, like in the measured AFM surface (fig. 4a). By taking the fast Fourier transform (FFT) of the surface corresponding to the image in frequency domain, these short range periodicities are particularly probed in this simulation and in the experiment. Figure 4b shows the FFT spectrum of the simulated random surface, which is a random distribution of data points in the frequency domain and is confined to a circular range of k-vectors, similar to the random photoinduced surface (fig. 2.b).

This 2D Fourier transform does not show any specific frequency distributions, characterized by two symmetrical spots generally attributed to a short range periodicity, confirming the absence of short range periodicities.

This figure is in agreement with figure 2b, proving the similarity between these two surfaces. Hence, the name of quasi random surface for the self-photostructuration of the azopolymer 1 is justified.

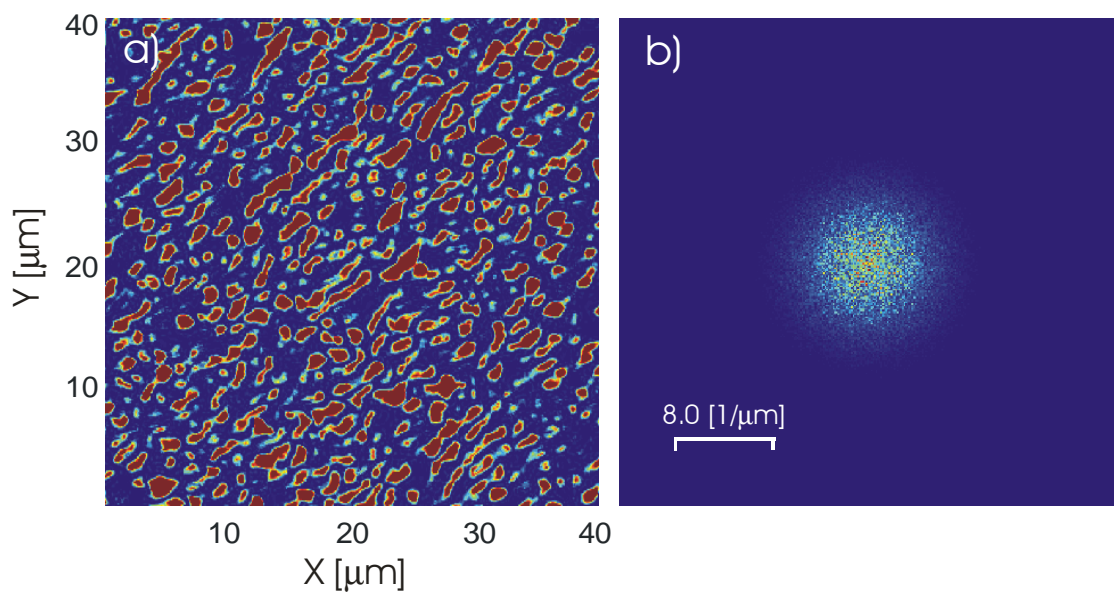

Fig.4. a) Simulation of a Gaussian surface considering the parameter of the photoinduced surface with the azopolymer 2, b) Fourier transform of the simulated surface.

### 3.2 Quasi random grating surface

The second experiment is done with the azopolymer 2. The figure 5a gives the topographical measurement of the surface after laser irradiation. The surface is spontaneously organized into sub-zones, oriented with different angles and periodic structures giving the aspect of quasi random periodic structurations. Two different types of subzones, where a periodic structure appears, are present. Each of these subzones contains a periodic structure with a pitch, defined as the distance between two maxima, of 0.51 ± 0.2 µm in the descending direction, or 0.35 ± 0.2 µm in the ascending direction. The subzones have an average length of 7.8 ± 0.3 µm, with a maximum width of 1.6 ± 0.2 µm. These zones inside the whole pattern appear to be superposed. Some parts of the sample show an overlapping between the randomly arranged sub-zones of gratings, where the periodic structures are photoinduced.

The self-organization of the thin film surface in multi gratings is the result of a mass transport due to *cis-trans* photoisomerization, as the light illuminates the surface [17]. In our experiment, the light illumination is done by a single Gaussian laser beam, without any structuration of impinging light. The long side chain of the second azopolymer with a high $T_g$, limits the process of molecular mass transports on the surface, in order to induce a large periodic structure by increasing the local viscosity. Another possible involved phenomenon could be the free volume size in the proximity of the azobenzene chromophore leading to opto-mechanical movements. The calculated molar volume of the sulfabenzamide azo derivative with the Gaussian09 package present in the azopolymer 1, is 293 cm$^3$/mol. While, this value for sulfadimethoxine azo derivative present in the azopolymer 2 is distinctly lower, 263 cm$^3$/mol (the molar volumes were calculated for *trans* isomers of model compounds). The lower molar volume of the photoisomerizable azobenzene derivative present in azopolymer 2 provides more free volume available for *trans-cis-trans* isomerization process. This may explain the higher photoisomerization and thermal relaxation rates observed for the azopolymer 2, favouring grating patterns compare to the azopolymer 1. Recently, we have presented results showing that, in the same condition for an azopolymer with a lower $T_g$, the self-organization takes place with features having a length of 300 ± 20 nm, revealing by the superposition of photoinduced periodic sub-structures [18]. An additional explanation could be based on the mechanism of phase separation of two coexistent immiscible phases in the polymer [19]. The phase could be attributed to the *trans* and *cis* domains. The local concentration of *trans* or *cis* isomers serves as a morphological order parameter controlling the undulation of the surface. Adjusting this local concentration with different contributions allows simulating different spontaneous surface patterns.

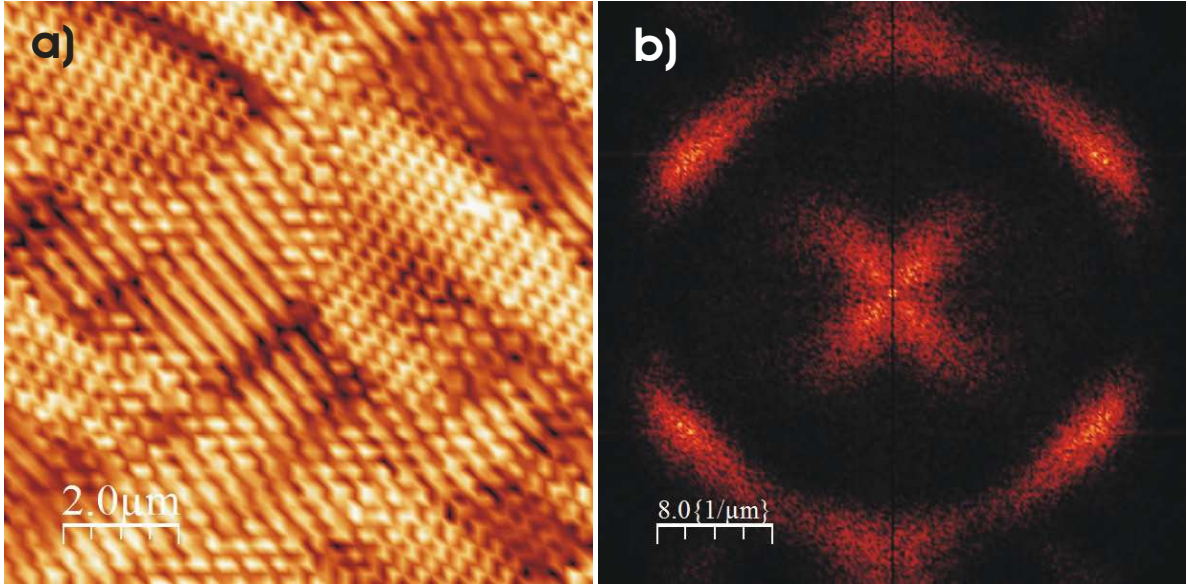

Fig.5. a) Surface pattern of the azopolymer 2 measured with an AFM in the contact mode, b) Fourier transform of the surface pattern.

The quasi random grating structuration is confirmed with the Fourier transform of the surface pattern. The figure 5b exhibits the Fourier spectrum of AFM measurements. The Fourier spectrum is confined to a circular range of k vectors, where the largest k-space distribution is due to the small structures confined in the multi gratings. The low k-values in the central part of the Fourier space are due to the multi gratings, layered in two different directions. We measured a maximal k-vector value of 17.2 µm$^{-1}$, larger than the measured values obtained with quasi random nanostructures on silicon materials [20]. The width of the k-vector ring is 4.5 µm$^{-1}$. Large k-vectors are a valuable property for light trapping applications. Indeed, the exit angle, $\theta_e$, of light diffracted by a grating surface is given by:

$$\sin \theta_e = q \lambda_r k_m \quad (6)$$

Where, q is the diffraction order, $\lambda_r$ is the reading wavelength and $k_m$ is the maximal value of k-vector in the Fourier transform of the surface pattern. If the exit angle, $\theta_e$, is greater than 90°, the light cannot emerge from the surface and is trapped inside the thin film. With a value of $k_m$ = 17.2 µm$^{-1}$, wavelengths as small as 360 nm do not exit the surface and are trapped inside the thin film, creating quasi-guided mode. This trapping effect is experimentally tested in section 3.3.

Consequently, we can say that the surface presented in figure 5a offers an interesting trade-off between periodic and random domains: the diffraction efficiency presents simultaneously a concentration of light into very high and very low orders.

A histogram of roughness analysis data, obtained by a measurement with an AFM, is shown in the figure 6a. The fit of the roughness can be done with two Gaussian curves and the histogram is in good agreement with these two curves.

The measured RMS roughness of the whole sample is σ = 20.6 nm. We distinguish an average roughness height of 62.5 nm and 91 nm for the first and the second average height respectively, given by the two roughness statistics fitted by a Gaussian curve.

The same roughness parameters were measured on different samples fabricated in the same conditions in a range of few percents confirming for this material the reproducibility of the spontaneous photoinduced patterns.

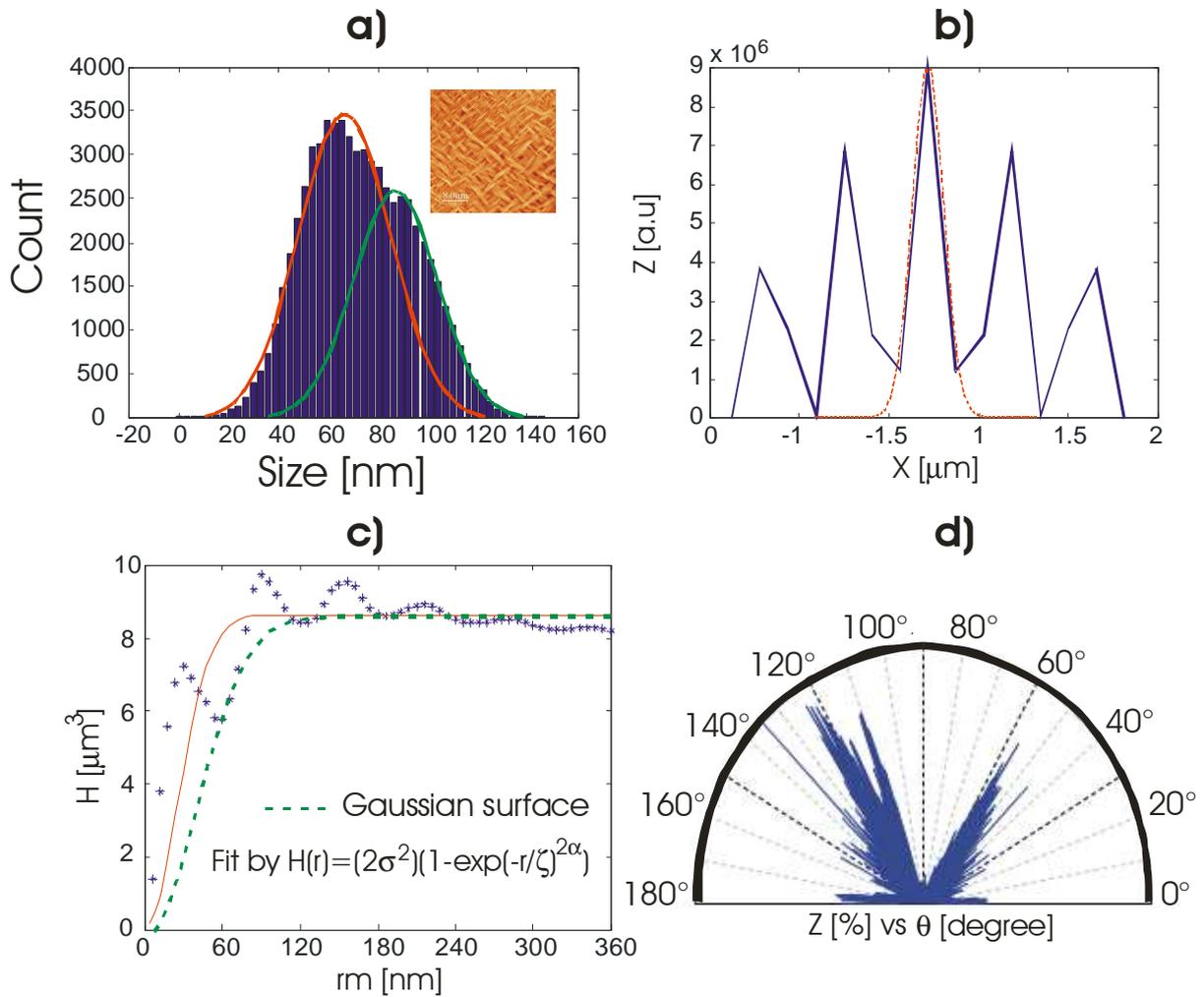

Fig.6. a) Histogram of the surface RMS roughness of the azopolymer 2 and its fit with a Gaussian curve, b) auto-correlation function of the pattern surface and comparison with a pure Gaussian surface (red), c) height–height correlation function (HHCF) and fit, d) directionality of the light emission from the surface.

The correlation function of surface heights (fig. 6b) gives a lateral correlation length of ξ = 98 ± 10 nm. This correlation function is not Gaussian. The different peaks are explained by periodicity of the pattern and we find the same correlation length in the other oscillations.

The ratio σ/ξ is 0.21. With this ratio, a large proportion of the surface becomes shadowed due to the topology for high angles of the incidence of light [21]. Any valleys on the surface start to be involved in multiple scattering events. So a part of primary reflections on the surface roughness give rise to secondary ones.

The height fluctuations occur at different length scales. The Hurst parameter corresponds to the frequency of height fluctuations. The surface roughness exponent parameter obtained from the fitting of the AFM scan is $\alpha = 0.63$, far from the value $\alpha = 1$, indicating a rough surface. The figure 6d shows the light directionality of the quasi-random gratings on the surface. We calculated an isotropy of 18.3 % in the emission of light from the surface. The directionality of the emission covers mostly four directions of the space. This anisotropic emission is due to the large random variation of micrometer length gratings on the surface, associated with a nano-patterning with a honeycomb form inside the gratings. However close to the surface the directionality of the emission is almost in the surface plane, allowing the reduction of the thickness of the trapping film. The surface texture model can be thought as the superposition of emission from gratings formed by slits, separated from one another with different distances [22].

### 3.3 Quasi guided modes into the thin film

In order to test the validity of light coupling through quasi-guided modes into the thin film, we have created four additional grating structures next to the quasi-random grating surface (figure 7a). First, the central pattern corresponding to quasi random gratings is spontaneously created by illuminating the azopolymer 2 with a single beam. Afterwards, four gratings are made with a laser interferences pattern, illuminating the sample through a pinhole. The sample was rotated in order to inscribe the 4 gratings, one by one, at the cardinal points, close to the central quasi random grating surface. The laser interferences pattern is obtained by separating the beam from a linearly polarized laser beam via a beam splitter and recombining them with two mirrors.

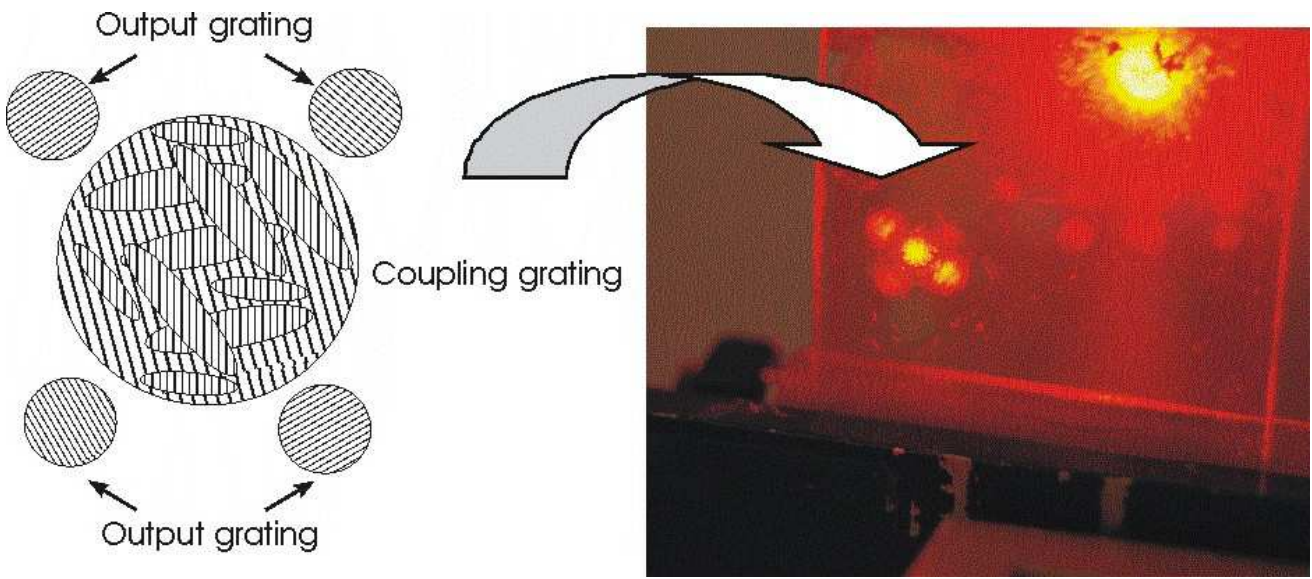

Fig.7. Four gratings were inscribed in the cardinal direction in order to show the mode coupling of the light in all the directions and the outcoupling light by the 4 gratings. The picture on the right shows that light (He-Ne laser) injected on the quasi random grating surface (coupling grating) is outcoupled by the output gratings. This confirms the existence of quasi guided modes into the thin film.

The positions of these four gratings match the positions of the diffracted spots given by the Fourier transform of the surface pattern (figure 5b). The pitch of these four gratings, measured from AFM images is $\Lambda = 0.9 \pm 0.05$ μm with an amplitude of $100 \pm 10$ nm. After inscription of these gratings, a He-Ne Laser is used to illuminate the central quasi-random grating pattern of the surface. We used a He-Ne laser in order to avoid possible photoinduced changes applied on the surface by the laser.

As it can be seen on figure 7b, the central pattern diffracts the incoming light with such big angles that a part of the incoming light cannot directly exit the film. However, the trapped light can excite the quasi-guided modes into the thin film. The guided energy is then outcoupled by the 4 satellite gratings. The variation of light intensity in the four gratings is due to amplitude inhomogeneities in the central pattern affecting the total repartition of the extracted light into the four gratings. Tilting the sample, with respect to the incident angle is important and can also affect the extracted light. The intensity of the light coming from the gratings can be controlled through changing the incident angle. In addition to applications for light harvesting, the quasi random-grating surface can also be seen as a four beam splitter for thin films.

### 3.4 Light trapping efficiency measurements

Due to the surface relief features, the surface-patterned azopolymer films can be used directly as a soft lithographic master to cast molds. We replicated the photoinduced surface of the chiral azopolymers 1 and 2 on a widely used elastomeric material, (poly(dimethylsiloxane (PDMS)). This material has a low interfacial free energy. The PDMS sample is largely transparent, since it has an optical absorption of 0.04 % and a transmission above 95 % in the visible region of the spectrum. The printed replicas showed the same random structures as the masters, along with the same relief depths. Figure 8 shows the photographic images of the stamps. Brilliant and iridescent colors, resulting from the diffraction of the surface gratings, can be observed for the replica of the quasi-random grating surface (sample A on figure 8).

Taking the plane PDMS surface as a reference, we measured the light extinction, induced by the patterned surface, for both materials. We found an increase of the light extinction ratio of 28% and 41% for the quasi Gaussian surface and the quasi-random grating surface, respectively. This light extinction is homogeneous over the visible spectrum, which is a valuable property for light harvesting. Since no absorption can occur inside the PDMS material, this extinction is due to light scattered by the random surface (for both surface) and light trapped into the thin film (only for the quasi random grating surface).

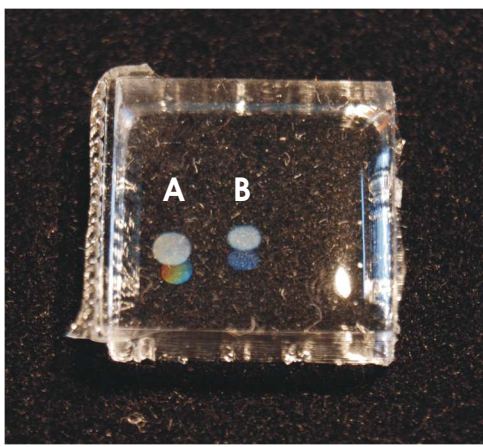
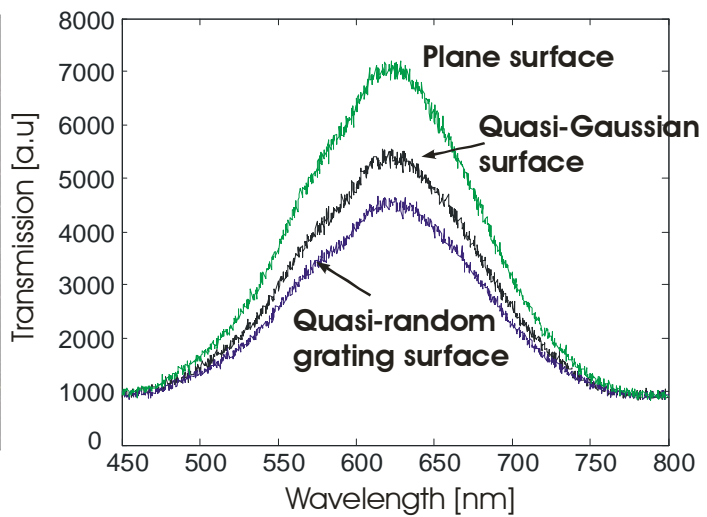

Fig.8. a) Replica modeling of the surface pattern of the azopolymer 1 and azopolymer 2 on an elastomeric material A and B respectively (A is the quasi-random grating surface). b) Transmission spectrum of the elastomeric surface with quasi-random gratings, a Gaussian surface and comparison with the reference plane surface.

Consequently, here light extinction has to be understood in the more general concept of light harvesting. Therefore, from the measurement we can say that the quasi-random grating surface is 20 % more efficient for light harvesting, than the quasi Gaussian surface. This effect is due to the random gratings associated with honey-comb structures inside the small gratings. We are similar or quasi better than the obtained results with photonics crystal [20, 23]. Let stress out again that in the quasi random grating structure, we combine the role of a random surface with the one of multi gratings to observe a large concentration of energy in higher orders, approaching the Lambertian limit. This replicated patterned surface can generate a good light trap.

## 4. CONCLUSION

In conclusion, we have presented a simple and cost effective way to prepare a quasi Gaussian surface and a quasi random grating surface using a self-photoinduced process on two chiral azopolymer thin films. The creation of a more interesting quasi random grating surface is possible, via the combination of a photoisomerization process associated with a good choice of the azopolymer material, in particular with the correct choice of the copolymer function and the grafted azobenzene. The quasi random grating surface has the same advantages as the quasi Gaussian surface for light scattering but, as it was experimentally shown, additionally it allows an induction of an isotropic mode coupling of light into thin film, increasing by 20% its efficiency for light harvesting.

We have highlighted the applicability of the quasi random grating surface for realistic devices. The replication of the surface pattern on a PDMS surface (a transparent and cost effective material) shows how easily this technique can be used to modify samples like OLEDs or solar cells for a better efficiency of light diffusion or light harvesting [24 - 27]. Furthermore, we note that the primary photoinduced surface can be erased for re-illumination and regeneration of a new photostructured surface. Finally, the chemical structure of the azobenzene grafted copolymer can be engineered, leading to new observational patterns.

**Acknowledgements**

We thank Romain Mallet (SCIAM – University of Angers) for the use of their equipment.
Calculations have been carried out using resources provided by Wroclaw Centre for Networking and Supercomputing (http://wcss.pl), grant No. 271